\pdfoutput=1
\documentclass[aps,pre,reprint,groupedaddress]{revtex4-2}

\usepackage{amsmath,amssymb}
\usepackage{graphicx}
\usepackage{xcolor}
\usepackage{hyperref}
\usepackage{orcidlink}
\usepackage{tikz}
\usetikzlibrary{quantikz2}
\providecommand{\orcid}[1]{\altaffiliation{ORCID: \href{https://orcid.org/#1}{#1}}}
\hypersetup{colorlinks=true,linkcolor=blue!60!black,citecolor=blue!60!black,urlcolor=blue!60!black}

\usepackage{eso-pic}
\begin{document}

\title{Fractal Period Locking and Action-Quantized Resonances
in the Open Quantum Baker Map}

\author{Leonardo Ermann}
\orcid{0000-0003-0964-3929}
\affiliation{DFTIFSC, GIyA, Comisi\'on Nacional de Energ\'ia 
At\'omica (CNEA) and CONICET, Buenos Aires, Argentina\\
ECyT,  Universidad Nacional de General San Mart\'in (UNSAM), 
B1653HIM San Mart\'in, Argentina}

\date{\today}
\begin{abstract}
We open the quantum baker map with a position window of width $w$ and
center $q_c$. Classical pruning tiles the $(q_c,w)$ plane into diamonds
where the $T$ leading resonances form root-of-unity multiplets. At
$w=1/2$ one periodic orbit survives for almost every $q_c$, and the
multiplets are quantized without free parameters: their modulus
$2^{-1/2}$ saturates the pressure bound and their phase is the orbit
action. The multiplets form cycling multicomponent cat states. Finally,
the opening reduces to one flag qubit per step, and on an IBM
superconducting processor the return probability already revives at the
period selected by the window.
\end{abstract}

\maketitle

Open quantum systems whose classical limit is chaotic have been far
less explored than their closed counterparts~\cite{altmann2013leaking}.
The lack of unitarity replaces energy levels by resonances. Their number
is governed by the fractal dimension of the classical repeller through
the fractal Weyl law~\cite{lu2003fractalweyl,nonnenmacher2007distribution,potzuweit2012weyl},
their decay rates are bounded by its topological
pressure~\cite{nonnenmacher2009decay,novaes2013resonances}, and their
long-lived eigenfunctions scar its shortest periodic
orbits~\cite{ermann2008periodic,ermann2009localization,nonnenmacher2007resonant,keating2006semiclassical}.
Resonances can be quantized semiclassically by periodic-orbit cycle
expansions~\cite{gaspard1989semiclassical,cvitanovic1989periodic}, and
approximated by scar functions built on short periodic orbits, whose
phases follow a Bohr-Sommerfeld-like rule~\cite{ermann2008periodic,novaes2009quantum,pedrosa2012short};
in the Walsh-quantized baker map the resonances are exactly organized in such
orbit multiplets~\cite{keating2008resonance}.
For a generic repeller, however, these constructions require resumming
exponentially many orbits, or a basis that grows with $N$. Therefore, an
exact rule assigning individual resonances to individual orbits is in
general out of reach.

The quantum baker map~\cite{balazs1989baker,saraceno1990classical,saraceno1994semiclassical}
is a simple model for this problem. Its classical counterpart is
piecewise linear and uniformly hyperbolic, and its quantization is built
from quantum Fourier transforms, so that it can be implemented
efficiently on a register of $n$
qubits~\cite{schack1998using,brun1999realizing,weinstein2002experimental}.
Opened by a fixed strip, it was used to establish the fractal Weyl law
and the scarring of resonances~\cite{nonnenmacher2007distribution,pedrosa2009distribution,ermann2009localization},
and its asymmetric, partially open version was recently shown to
interpolate between Poisson and Ginibre spacing
statistics~\cite{ermann2026complexspacing}.

In this Letter we let the opening slide, and study the leading
resonances of the symmetric quantum baker map over the whole plane of
window centers and widths. We show that the classical pruning tiles
this plane into diamonds bounded by border-collision lines. Inside each
diamond a single orbit of period $T$ survives, and the $T$ leading
resonances lock into a root-of-unity multiplet, the scar multiplet of
that orbit. When the window is a half circle, one orbit survives for
almost every window position, and its period follows a devil's
staircase, an effect that we call fractal period locking. There the
multiplets are quantized without free parameters. Their modulus
saturates the topological-pressure bound at $2^{-1/2}$, and their phase
is the classical action of the orbit, which makes it periodic in the
number of qubits. Moreover, the multiplets recombine into
multicomponent cat states that cycle around the orbit. Finally, the
opening reduces to a single flag qubit per step, and period, Lyapunov
exponent and action can be measured on a quantum processor in a few
steps; pilot runs on a superconducting processor already resolve the
period locking.

\begin{figure*}
  \centering
  \includegraphics[width=\textwidth]{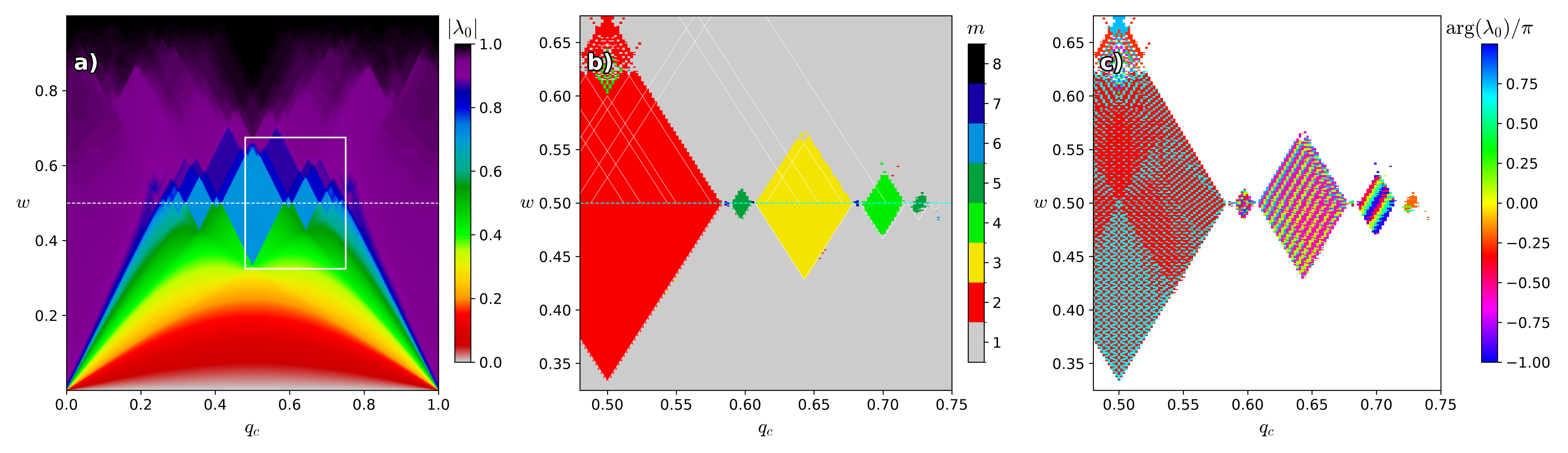}
  \caption{Leading resonances over the plane of window centers and
  widths at $n=10$ ($N=1024$); the dashed line marks $w=1/2$.
  (a)~Modulus $|\lambda_0|$. All diamonds share
  $|\lambda_0|\simeq2^{-1/2}$, except the cones of the fixed point
  $q=0$ near the edges ($|\lambda_0|\simeq0.9$--$0.95$). Between these cones
  and below the diamonds no orbit survives and $|\lambda_0|$ decreases
  to zero with $w$; positive-entropy repellers, with $|\lambda_0|$
  approaching 1, appear only for $w>1/2$. The rectangle
  marks the region shown in (b) and (c). (b)~Multiplet size $m$, with
  the border-collision lines of Eq.~\eqref{eq:diamond} (white).
  (c)~Phase $\arg(\lambda_0)/\pi$ where $m>1$: inside each diamond
  $\arg\lambda_0$ takes one of the $T$ values
  $(\Phi_{\mathcal{O}}+2\pi k)/T$ of Eq.~\eqref{eq:multiplet}, hence the
  checkerboard pattern.}
  \label{fig:pruning}
\end{figure*}

\textit{Model.}---The closed map acts on $\mathcal{H}=\mathbb{C}^N$,
$N=2^n$, as
\begin{equation}
  \label{eq:closed_U}
  U = G_N^{\dagger}\begin{pmatrix} G_{N/2} & 0 \\ 0 & G_{N/2}\end{pmatrix},
  \quad
  (G_N)_{jk}=\frac{e^{-2\pi i (j+\frac12)(k+\frac12)/N}}{\sqrt N},
\end{equation}
where $G_{N/2}$ is defined in the same way with $N\to N/2$. This is the Balazs-Voros-Saraceno
quantization~\cite{balazs1989baker,saraceno1990classical} of the baker
map $(q,p)\mapsto(2q-\epsilon,\,(p+\epsilon)/2)$, $\epsilon=\lfloor 2q
\rfloor$, with effective Planck constant $h=1/N$ on the position grid
$q_j=(j+\frac12)/N$. The classical map is area preserving and uniformly
hyperbolic with Lyapunov exponent $\lambda_{\rm L}=\ln 2$, and its stable and
unstable foliations are $\{q={\rm const}\}$ and $\{p={\rm const}\}$. At
every step we keep only the population inside the position window
$W(q_c,w)=[q_c-w/2,\,q_c+w/2)$ (mod 1),
\begin{equation}
  \label{eq:propagator}
  \mathcal{U} \;=\; U\,\Pi_W,
\end{equation}
where $\Pi_W$ projects onto the ${\rm round}(wN)$ grid sites of $W$: a
state is first projected, and the survivors are then propagated by $U$.
Removing instead a hole of width $1-w$ centered at $q_c+1/2$ gives the
same operator, so that Eq.~\eqref{eq:propagator} describes equally a
survival window of width $w$ and an escape hole of width $1-w$. The eigenvalues $\lambda_k$ of
$\mathcal{U}$, ordered by decreasing modulus, are the resonances of the
open map~\cite{nonnenmacher2011openmaps,novaes2013resonances}, and
$|\lambda_0|^2$ gives the asymptotic decay of the surviving population
per step. The reflection $(q,p)\to(1-q,1-p)$ makes the spectrum exactly
symmetric under $q_c\to 1-q_c$.

\textit{Classical pruning.}---Classically, $\mathcal{U}$ reduces to the
doubling map $q\mapsto 2q$ mod 1 restricted to $W$. A primitive period-$T$
orbit $\mathcal{O}$ survives if and only if all its points lie in $W$,
that is,
\begin{equation}
  \label{eq:diamond}
  w \;\ge\; \Delta_{\mathcal{O}} + 2\,|q_c-\bar q_{\mathcal{O}}|,
\end{equation}
where $\Delta_{\mathcal{O}}$ is the length of the shortest arc covering
$\mathcal{O}$ and $\bar q_{\mathcal{O}}$ is its center. Each orbit thus
survives inside a cone of universal slopes $\pm2$ with apex
$(\bar q_{\mathcal{O}},\Delta_{\mathcal{O}})$, bounded by border-collision
lines~\cite{nusse1992border} where a window edge grazes an orbit point.
Where a single orbit survives, its cone is closed from above by the
cones of the next orbits entering $W$, and the plane is tiled by
diamonds, in close analogy with the mode locking of circle
maps~\cite{jensen1983complete}. For instance, the period-2 orbit
$\{1/3,2/3\}$ produces the central diamond of Fig.~\ref{fig:pruning},
with lower apex $(1/2,1/3)$, which is closed at $w=3/5$ when the
period-4 orbit $\{1/5,2/5,4/5,3/5\}$ enters the window. In
Fig.~\ref{fig:pruning}(b) we show that the quantum spectrum follows this
tiling closely. Inside each diamond the leading eigenvalues form a
multiplet of size $m=T$, quasidegenerate in modulus and equally spaced
in phase, which we detect from the spectrum alone~\cite{SM}. Where two
isolated orbits coexist, as at $q_c=1/2$ for $3/5\le w\lesssim0.65$, the
spectrum contains both multiplets~\cite{SM}. Notice that all diamonds
share the same leading modulus $|\lambda_0|\simeq0.71$
[Fig.~\ref{fig:pruning}(a)]. The only exception is the cone of the fixed
point $q=p=0$, near the edges of Fig.~\ref{fig:pruning}(a), where
$|\lambda_0|\simeq0.9$--$0.95$ at $n=10$, slowly increasing with $N$
($0.971$ at $n=26$~\cite{SM}). This orbit lies at the corner of the torus, on
the discontinuity of the map, where the semiclassics of the quantum
baker map is known to be anomalous~\cite{lakshminarayan1995unusual,toscano1997boundary}.
The points $\nu/(2^T-1)$ of all orbits with $T\ge2$ stay away from
it~\cite{SM}.

\begin{figure}
  \centering
  \includegraphics[width=\columnwidth]{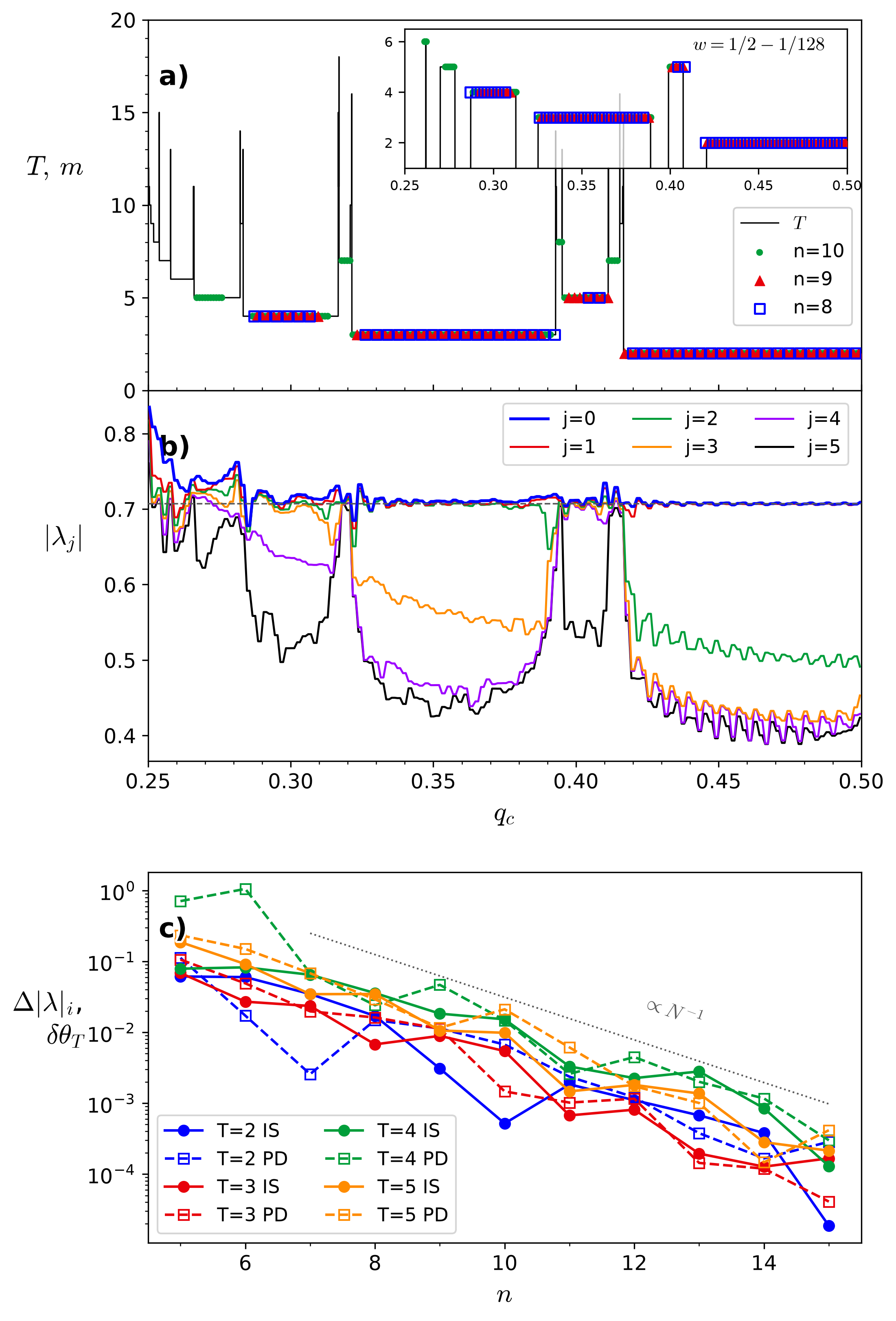}
  \caption{Period locking at $w=1/2$ for $1/4\le q_c\le1/2$ (for
  $q_c<1/4$ the fixed point survives, and $q_c\to1-q_c$ is an exact
  symmetry). (a)~Classical period $T$ (black) and multiplet size $m$ for
  $n=8$, $9$, $10$; inset: $w=1/2-1/128$. (b)~Moduli of the six leading
  eigenvalues at $n=10$, labeled by their rank $j=0,\dots,5$ in
  decreasing modulus at each $q_c$ (not by the index $k$ of
  Eq.~\eqref{eq:multiplet}); $j=0$ is drawn on top. On each plateau the
  multiplet sits at $|\lambda_j|\simeq2^{-1/2}$ (dashed line). (c)~Intra-band spread
  $\Delta|\lambda|_i$ (IS, full circles) and phase deviation
  $\delta\theta_T$ (PD, open squares) versus $n$, at $q_c=0.50$, $0.35$,
  $0.30$ and $25/62$ inside the $T=2,\dots,5$ plateaus; dotted line:
  $\propto N^{-1}$.}
  \label{fig:staircase}
\end{figure}

On the line $w=1/2$ the window is a half circle, and the surviving set
is either a single periodic orbit or a Sturmian Cantor set. Its rotation
number follows a devil's staircase as $q_c$ is
varied~\cite{bullett1994ordered}, as for the doubling map with a
hole~\cite{sidorov2014supercritical,glendinning2015doubling}. For a rotation number $\rho=r/T$, with $r$ and $T$ coprime, the surviving orbit has span
$\Delta_{\mathcal{O}}=\frac12-\frac{1}{2(2^T-1)}$, so that by
Eq.~\eqref{eq:diamond} it survives alone on a plateau of width
$\delta q_c=1/[2(2^T-1)]$. Summing over the $\varphi(T)$ rotation numbers
of each period, the plateaus tile the circle exactly,
$\sum_{T\geq1}\varphi(T)/[2(2^T-1)]=1$, as follows from the Lambert
series $\sum_T\varphi(T)x^T/(1-x^T)=x/(1-x)^2$ at
$x=1/2$~\cite{SM}. Thus a single periodic orbit survives for almost
every $q_c$. In fact, since any arc of length $w\le1/2$ lies inside a
half circle, the repeller has zero topological entropy over the whole
lower half $w\le1/2$ of the plane, and positive-entropy repellers appear
only for $w>1/2$ [Fig.~\ref{fig:pruning}(a)]. Figure~\ref{fig:staircase}(a)
shows that the quantum multiplet size reproduces the classical
staircase.

\begin{figure*}
  \centering
  \includegraphics[width=0.72\textwidth]{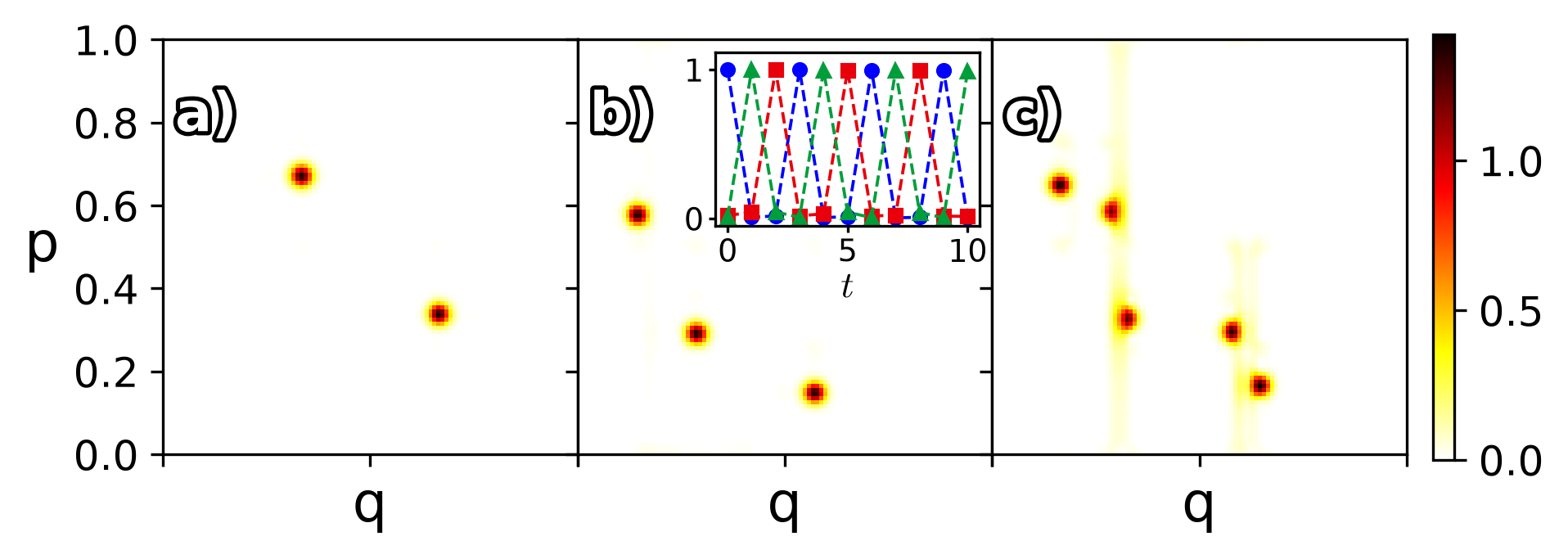}
  \caption{Biorthogonal distribution $\sum_k|h_k(q,p)|$ of the multiplet
  at $w=1/2$, $n=9$, for (a)~$T=2$ ($q_c=1/2$), (b)~$T=3$ ($q_c=0.35$)
  and (c)~$T=5$ ($q_c=25/62$, plateau center). It is concentrated on the
  orbit points, e.g., $(q,p)=(1/7,4/7),(2/7,2/7),(4/7,1/7)$ for $T=3$, and
  its maximum exceeds 1 because $\psi_k^{R,L}$ are biorthogonally rather
  than individually normalized, a Petermann-type excess due to their
  nonorthogonality~\cite{chalker1998eigenvector}. Inset: cycling fidelity
  $\mathcal{F}_j(t)$ for the plateau of (b), with $j=0$, $1$, $2$ in
  blue, red and green.}
  \label{fig:cat}
\end{figure*}

\textit{Semiclassical quantization.}---Inside a plateau with $T\ge2$,
$\mathcal{U}$ maps states launched from the $T$ orbit points onto each
other cyclically. At each step the amplitude is multiplied by
$\Lambda^{-1/2}=2^{-1/2}$, with $\Lambda=2$ the stretching factor of the
unstable direction, and acquires the phase of the generating function of
the map~\cite{saraceno1994semiclassical}. Diagonalizing this $T\times T$
cyclic block gives the multiplet
\begin{equation}
  \label{eq:multiplet}
  \lambda_k \simeq 2^{-1/2}\, e^{i(\Phi_{\mathcal{O}}+2\pi k)/T},
  \qquad
  \Phi_{\mathcal{O}} = -2\pi N S_{\mathcal{O}} \;\;({\rm mod}\;2\pi),
\end{equation}
$k=0,\dots,T-1$, where $S_{\mathcal{O}}=\sum_t
[F(q_t,p_{t+1})-q_{t+1}p_{t+1}]$ is the action of the orbit, computed
with the generating function $F(q,p')=2qp'-\epsilon\,(q+p')$ of the
baker map, and no Maslov phase appears. Since $q_{t+1}=2q_t-\epsilon_t$,
each step contributes exactly $-\epsilon_tq_t$, independently of the
momenta, so that the action is minus the sum of the orbit points in the
right half of the square, $S_{\mathcal{O}}=-\sum_{t:\,\epsilon_t=1}q_t$. Both factors are free of
adjustable parameters. Equation~\eqref{eq:multiplet} is the map
counterpart of the quantization of resonances generated by a single
closed hyperbolic trajectory~\cite{gerard1987semiclassical}. Its phase is the
Bohr-Sommerfeld-like exponent $(NS_{\mathcal{O}}+k)/T$ of the scar functions of
Refs.~\cite{ermann2008periodic,novaes2009quantum}, up to the sign
convention of $S_{\mathcal{O}}$, and its root-of-unity structure is the exact one of
the Walsh-quantized baker map~\cite{keating2008resonance}. The new
ingredient is that the opening leaves one orbit alone in the repeller for
almost every $q_c$. Therefore, the orbit modes are themselves the
resonances, without interference from other orbits and without a
scar-matrix diagonalization. The modulus $2^{-1/2}=e^{-\lambda_{\rm L}/2}$ saturates
the bound $|\lambda_0|\le e^{P(1/2)}$ set by the topological pressure
$P(s)$ of the repeller~\cite{nonnenmacher2009decay,novaes2013resonances},
since a single orbit has zero entropy and $P(1/2)=-\lambda_{\rm L}/2$.
Equivalently, the quantum decay $|\lambda_0|^2=1/2$ per step coincides
with the classical escape rate $e^{-\lambda_{\rm L}}$ of a zero-entropy
repeller~\cite{kantz1985repellers}. In Fig.~\ref{fig:staircase}(b) the
$m$ leading eigenvalues sit at $2^{-1/2}$ on each plateau, and at $n=15$ ($N=32768$) the moduli of the
$T=2,\dots,5$ multiplets equal $2^{-1/2}$ within
$1.1\times10^{-4}$~\cite{SM}. On the other hand, the
action is a rational number with denominator $2^T-1$: for the orbits
$\{1/3,2/3\}$, $\{1,2,4\}/7$, $\{1,2,4,8\}/15$ and $\{5,10,20,9,18\}/31$
of the $T=2,\dots,5$ plateaus we find $S_{\mathcal{O}}=-2/3$, $-4/7$,
$-8/15$ and $-38/31$. As a consequence, the phase follows the doubling
map itself as qubits are added,
$\Phi_{\mathcal{O}}(n+1)=2\Phi_{\mathcal{O}}(n)$ mod $2\pi$, and is
periodic in $n$ with period $T$. For the period-2 orbit it
alternates between $\pm2\pi/3$ with the parity of $n$. Equation~\eqref{eq:multiplet} holds equally for any even $N$~\cite{SM}. At the plateau
centers the measured phases $\arg\lambda_k^T$ follow
Eq.~\eqref{eq:multiplet} within $3.3\times10^{-3}$ for all four orbits
already at $n=12$, and within $6\times10^{-4}$ at $n=15$~\cite{SM}. Moreover, at fixed $n$ the phase is the
same over the whole diamond: over the $2\times10^4$ grid points of
Fig.~\ref{fig:pruning} with $m=2,3,4$, it agrees with
Eq.~\eqref{eq:multiplet} within $1.2\times10^{-3}$ of a turn~\cite{SM}.

The quantum staircase resolves plateaus far narrower than a minimal wave
packet. At $n=10$ the plateaus of rotation numbers $2/7$, $3/8$ and
$3/7$, only four, two and four grid sites wide, are already detected
[Fig.~\ref{fig:staircase}(a)], while a minimal packet spans
$\sqrt{N}=32$ sites. This can be understood as follows. The long-lived
left eigenvectors concentrate on the stable foliation through the orbit
points~\cite{keating2006semiclassical}, the vertical lines $q=q_t$, and
are therefore sharp in position. At $n=9$, about $90\%$ or more of their
weight lies within one grid site of the orbit points~\cite{SM}, whereas the right
eigenvectors spread along the unstable foliation over the whole window. As a consequence, plateaus are resolved down to a width of about two grid
sites, $N/[2(2^T-1)]\gtrsim2$, that is, for periods up to
$T\simeq n-2=\log_2N-2$, twice the Ehrenfest time $t_E\simeq n/2$, for
orbits that stay away from the corner fixed point~\cite{SM}.
Since removing grid sites $\delta\Pi$ from $W$ changes $\lambda_k$ by $-\lambda_k\langle\psi^L_k|\delta\Pi|\psi^R_k\rangle$ to first order (with $\langle\psi^L_k|\psi^R_k\rangle=1$),
the multiplet ignores the window edges until they come within a few
sites of an orbit point. As a result, Eq.~\eqref{eq:multiplet} holds
down to the lattice scale. At $n=10$ the multiplet of the $\rho=3/8$
plateau, whose orbit points lie one grid site from the window edges,
has a mean modulus within $3\times10^{-4}$ of $2^{-1/2}$ and a phase
within $0.015$ rad of Eq.~\eqref{eq:multiplet}~\cite{SM}. The remaining finite-$N$ corrections, the
intra-band spread $\Delta|\lambda|_i=|\lambda_0|-|\lambda_{T-1}|$ (IS)
and the largest deviation $\delta\theta_T$ of the phases from a regular
$T$-gon (PD), decay approximately as $N^{-1}$, i.e., linearly in $h$
[Fig.~\ref{fig:staircase}(c)].

\begin{figure}
  \raggedright{\bfseries\sffamily a)}\\[-4pt]
  \centering
  \resizebox{0.97\columnwidth}{!}{%
  \begin{quantikz}[row sep={0.75cm,between origins}, column sep=0.32cm]
  \lstick{flag $\ket{0}$} & & & \gate[3]{C_{q_c}}\gategroup[3,steps=4,style={dashed,rounded corners,inner xsep=3pt,inner ysep=4pt},label style={label position=below,anchor=north,yshift=-0.2cm}]{one step of $\mathcal{U}$, repeated $t$ times} & \meter{} & \setwiretype{n} & & \\
  \lstick{$j_{n-1}$} & & & & & & \gate[2]{G_{N}^{\dagger}} & \\
  \lstick{$j_{n-2}\cdots j_0$} & \qwbundle{n-1} & & & & \gate{G_{N/2}} & &
  \end{quantikz}}\\[2pt]
  \includegraphics[width=0.92\columnwidth]{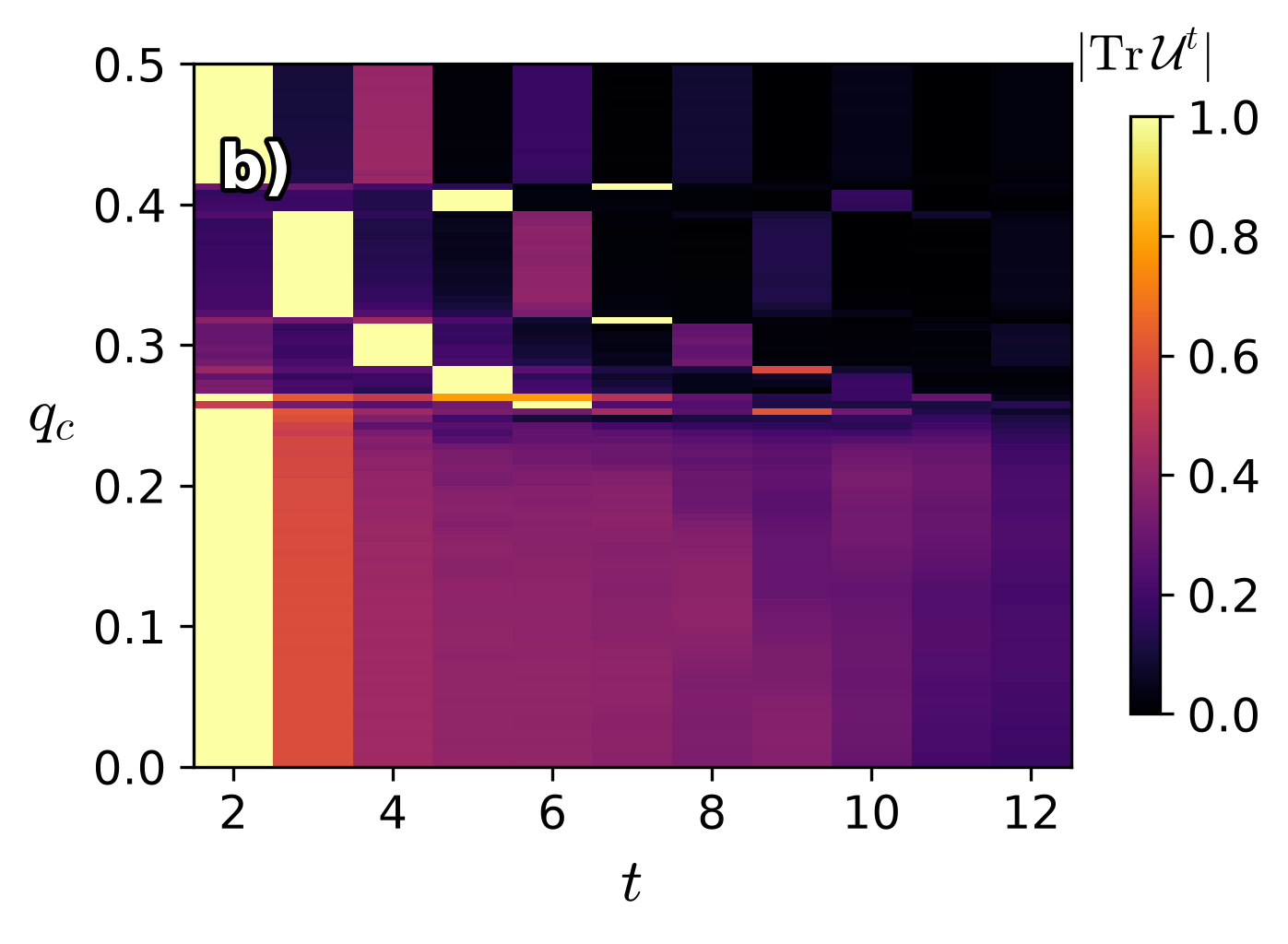}\\[2pt]
  \raggedright{\bfseries\sffamily c)}\\[-4pt]
  \centering
  \includegraphics[width=0.93\columnwidth]{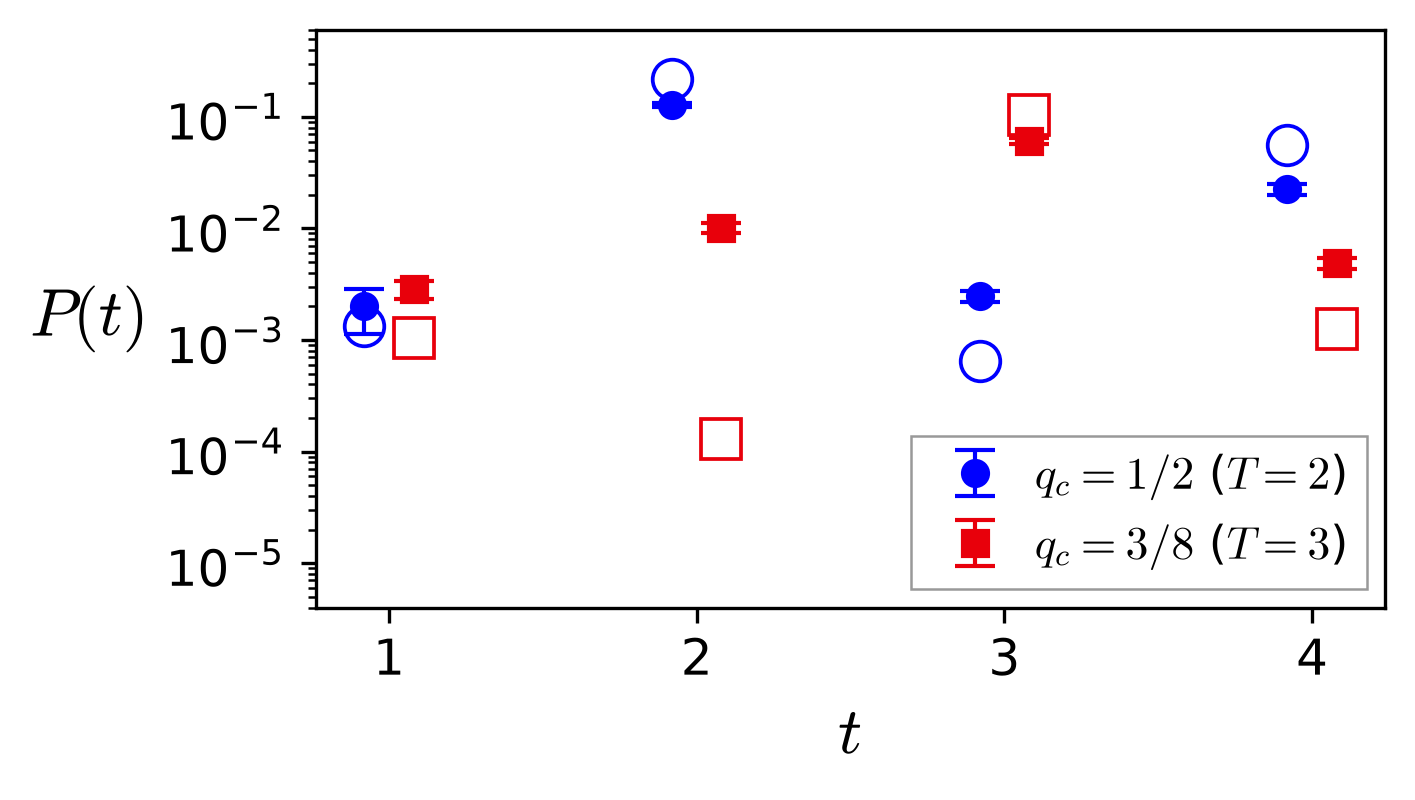}
  \caption{(a)~One step of $\mathcal{U}$ at $w=1/2$: the comparator
  $C_{q_c}$ writes on a flag qubit whether the position register lies
  outside $W(q_c,1/2)$, the flag is measured mid-circuit (runs are kept
  when it reads 0) and reset, or written on a fresh ancilla measured at
  the end, and the closed map
  $U=G_N^\dagger(\mathbb{I}\otimes G_{N/2})$ is
  applied; for $q_c=1/4$, $C_{q_c}$ is a CNOT from the most significant
  qubit $j_{n-1}$ of $j=\sum_bj_b2^b$. (b)~Blind spectroscopy at $n=10$:
  $|\mathrm{Tr}\,\mathcal{U}^t|$ for $t=2,\dots,12$, normalized for each
  $q_c$ by its maximum, revives at multiples of the period and
  reconstructs the staircase of Fig.~\ref{fig:staircase}(a) without
  orbit information. (c)~Return probability $P(t)$ of the grid state
  closest to an orbit point at $n=5$, $w=1/2$, for $q_c=1/2$ ($T=2$,
  circles) and $q_c=3/8$ ($T=3$, squares), measured on
  \texttt{ibm\_kingston} (full symbols: mean and standard deviation over
  five runs of $6000$ shots) and exact (open symbols).}
  \label{fig:circuit}
\end{figure}

\textit{Cycling cat states.}---The $T$ right eigenvectors
$|\psi_k^R\rangle$ of a plateau are each spread over the whole orbit,
but the biorthogonal phase-space distribution
$h_k(q,p)=\langle q,p|\psi_k^R\rangle\langle\psi_k^L|q,p\rangle$~\cite{ermann2009localization},
which combines both foliations, localizes on the $T$ orbit points
[Fig.~\ref{fig:cat}(a)-(c)]. Recombining the multiplet by a discrete
Fourier transform,
\begin{equation}
  \label{eq:localized_packets}
  |W_j\rangle = \frac{1}{\sqrt{T}}\sum_{k=0}^{T-1}
  e^{-2\pi i jk/T}\, e^{-i\phi_k}\,|\psi_k^R\rangle,
  \qquad j=0,\dots,T-1,
\end{equation}
with gauge phases $\phi_k$ chosen to interfere constructively at a
reference site, gives $T$ states that are extended in position and
concentrated in momentum near the unstable lines $p=p_t$ through the orbit points, strongly for $T=2$ and more loosely for larger $T$. The
same recombination of the left eigenvectors gives their duals, which
are concentrated in position near the stable lines $q=q_t$, and the
biorthogonal product of the two localizes on the orbit
points~\cite{SM}. They are the open-map counterparts of the periodic-orbit modes and scar functions of the closed and open baker map~\cite{ermann2008periodic,novaes2009quantum}.
Equation~\eqref{eq:multiplet} then implies
\begin{equation}
  \label{eq:cyclic_evolution}
  \mathcal{U}\,|W_j\rangle \simeq 2^{-1/2}\,
  e^{i\Phi_{\mathcal{O}}/T}\, |W_{j-1\bmod T}\rangle,
\end{equation}
so that the open map shifts these states cyclically along the orbit,
with no dispersion. The inset of Fig.~\ref{fig:cat}(b) confirms this: the normalized
cycling fidelity
$\mathcal{F}_j(t)=|\langle W_j|\mathcal{U}^t|W_0\rangle|^2/
(\|W_j\|^2\|\mathcal{U}^tW_0\|^2)$ returns to unity at $t\equiv-j$ (mod
$T$), with peaks above $0.989$ over $10$ steps at $n=9$, and stays
below $0.05$ otherwise. Any superposition of the $|W_j\rangle$ is thus a
$T$-component cat state that keeps its shape under the open dynamics.
Conditioned on survival, it cycles around the orbit. In the closed map,
on the other hand, each step splits a line $p=p_t$ into two parallel
lines, and the state fills the torus after about $n=\log_2N$ steps. Its
norm decreases by the factor $2^{-T}$ per period~\cite{SM}, which is the
price of a stabilization produced by the projection itself. In contrast to
driven-dissipative cat qubits, where a two-component manifold is fixed
by an engineered drive~\cite{mirrahimi2014dynamically}, here the number
of components is selected by the classical pruning and retuned by
sliding $q_c$ along the staircase.

\textit{Circuit implementation.}---The closed map~\eqref{eq:closed_U}
reads $U=G_N^\dagger(\mathbb{I}\otimes G_{N/2})$, where the antiperiodic
transforms $G_N$ and $G_{N/2}$ are quantum Fourier transforms dressed by single-qubit
phases, as already exploited in NMR
implementations~\cite{schack1998using,brun1999realizing,weinstein2002experimental}.
At $w=1/2$ the projector $\Pi_W$ tests whether the position register
lies in a half circle, that is, the constant comparison
$(j-j_0)\bmod N<N/2$ with $j_0=N(q_c-1/4)$. This comparison can be
written on a flag qubit with $O(n)$ gates~\cite{cuccaro2004adder}, and
for $q_c=1/4$ it reduces to reading the most significant qubit
[Fig.~\ref{fig:circuit}(a)]. In a register shifted by $j_0$ it is that
qubit for every $q_c$, at the cost of one constant adder per
step~\cite{SM}. One step of $\mathcal{U}$ is thus a
mid-circuit measurement of the flag followed by $U$. Runs that survive
$t$ steps implement $\mathcal{U}^t$, and the complementary record
samples the window $W(q_c+\frac12,\frac12)$ in the same runs.
Equivalently, the flag of each step can be written on a fresh ancilla
that is measured only at the end~\cite{nielsen2010quantum}, so that
neither mid-circuit measurements nor resets are needed. The open
baker map at $w=1/2$ is then a simple monitored
circuit~\cite{li2019measurement,skinner2019measurement}, with one qubit
measured at every step instead of at random. Mid-circuit measurements
and resets in chaotic-map circuits have already been run on up to 100
qubits~\cite{pokharel2026order}. The postselection overhead, a factor
$|\lambda_0|^{-2}=2$ per step, does not grow with $n$, in contrast with
the cost of reproducing a measurement record in a generic monitored
circuit, which is exponential in the total number of measurements.
Equation~\eqref{eq:multiplet} can thus be tested directly. For a minimal
coherent state $|q_t,p_t\rangle$ prepared at a point of the
orbit~\cite{kitaev2008wavefunction}, known classically from its
symbolic word, the return amplitude
$A(t)=\langle q_t,p_t|\mathcal{U}^t|q_t,p_t\rangle$ vanishes unless
$t=sT$, its modulus decreases by $2^{-T/2}$ between successive
revivals, and its phase at $t=T$ equals $\Phi_{\mathcal{O}}$, within
$2\times10^{-3}$ at $n=10$~\cite{SM}. Since
$|A(T)|=0.69$, $0.50$, $0.35$ and $0.25$ for $T=2,\dots,5$, on an
ideal device the period and the Lyapunov exponent follow from $O(10^2)$
shots, and the action, within $0.1$~rad, from a Hadamard test with
$O(10^3)$ shots, independently of $N$. Even without any orbit
information, the normalized trace $\mathrm{Tr}\,\mathcal{U}^t/N$, which
a one-clean-qubit estimator~\cite{knill1998power} with postselected
flags provides~\cite{SM}, revives at $t=sT(q_c)$ and reconstructs the
staircase [Fig.~\ref{fig:circuit}(b)]. The cost is of the order of
$N^2$ repetitions, since $|\mathrm{Tr}\,\mathcal{U}^T|$ remains of order
unity.

\textit{Pilot runs.}---The main limitation is thus the depth, as for
the few-qubit chaotic-map circuits already run on present
processors~\cite{porter2025impact}. We ran the protocol on the IBM
superconducting processor
\texttt{ibm\_kingston}~\cite{ibmquantum}, with the flags written on
fresh ancillas and the register prepared in the grid state closest to
an orbit point~\cite{SM}. At $n=5$ one step of $\mathcal{U}$ costs 65--71
two-qubit gates after routing. Figure~\ref{fig:circuit}(c) shows the
period locking: for $q_c=1/2$ the return probability peaks at $t=2$,
and for $q_c=3/8$ at $t=3$, about $50$ and $6$ times above the largest
off-period value. The peaks are reduced by noise, $P(2)=0.128\pm0.005$
against $0.220$, but their ratio, $P(3)/P(2)=0.48\pm0.04$ over three
sessions, agrees with the exact value $0.468$. Deeper circuits, such as
the $T=4$ revival at $n=6$, were not resolved~\cite{SM}. In the Walsh-quantized baker
map~\cite{schack2000shifts,keating2008resonance}, where the multiplet
law is exact and one step is a single Hadamard gate followed by a
relabelling of the qubits, the same protocol reaches $n=12$
($N=4096$). There the return probability peaks at $t=T$ for $T=2$ and
$3$, with no return at all at off-period times up to $n=10$. The decay
between revivals gives $T^{-1}\ln[P(T)/P(2T)]=0.70$--$0.82$ for
$n=6$--$12$, against the Lyapunov exponent $\ln2=0.693$, with an upward
bias due to noise. A Hadamard test~\cite{cleve1998revisited} resolves the sign of $A(2)$,
which alternates with the parity of $n$ as the action phase
predicts~\cite{SM}. The action phase of the Fourier multiplets, which
requires controlled steps, remains to be measured.

\textit{Conclusions.}---In this work we have shown that sliding the opening of the
quantum baker map organizes its leading resonances into a diamond tiling
of root-of-unity multiplets, one per surviving periodic orbit. For
survival windows no larger than a half circle the repeller has zero
entropy. At $w=1/2$ a single orbit survives for almost every window
position, and the multiplets are quantized without free parameters:
their modulus saturates the pressure bound at
$2^{-1/2}$, and their phase is the classical action of the orbit,
periodic in the number of qubits. The quantum staircase resolves
plateaus down to the lattice scale, because the left eigenvectors live
on the stable foliation, and the multiplets recombine into cycling
multicomponent cat states. Finally, at $w=1/2$ the opening is a single
flag qubit per step. Period, Lyapunov exponent and action can then be
read off a few-step circuit, and pilot runs on a superconducting
processor already resolve the period locking.

Of course, the universality of the modulus relies on the uniform
hyperbolicity of the baker map, and fails for the fixed point lying on
its discontinuity. For maps with nonuniform stretching the
multiplet modulus is set instead by the stability of each orbit,
$|\lambda_k|\simeq|\Lambda_{\mathcal{O}}|^{-1/(2T)}$, so that $|\lambda_0|$
itself follows the staircase. In the asymmetric baker map of
Ref.~\cite{ermann2026complexspacing} at $w=1/2$, $n=10$, we find
$|\lambda_0|=0.727$ and $0.690$ for single surviving orbits of period 3
and 2, against $0.727$ and $0.687$~\cite{SM}. The same test applies to
the generalized baker maps of Ref.~\cite{ermann2006generalized}, which
share the circuit structure of $U$, and to the quantized standard and
sawtooth maps. Near the accumulation points of the
staircase, where $T$ diverges, the finite-$N$ spectrum departs from
Eq.~\eqref{eq:multiplet} [Fig.~\ref{fig:staircase}(b)]. A first estimate
with position dephasing shows that the coherence of the cat states
degrades faster for larger $T$~\cite{SM}. On present superconducting
hardware, gate and readout noise reduce the revivals to $55$--$90\%$ of
their exact values and bias the decay rate upward by up to $20\%$.
Noisy simulations of the $n=6$ circuits show that two-qubit gate
infidelities below $10^{-3}$ with all-to-all connectivity, already
reached by trapped-ion processors~\cite{ransford2026helios}, give the
action phase of the $T=2$ and $T=3$ multiplets within $0.1$~rad with
about $10^3$ and $10^4$ shots, and the Lyapunov exponent within
$5\%$~\cite{SM}.
The next step is the measurement of the action phase on such a
device.

\begin{acknowledgments}
The author thanks G.~G. Carlo and M. Saraceno for useful discussions.
This work was supported by CONICET under Grant No.\ PIP 2022-2024 GI --
11220210100208CO. The author acknowledges the use of IBM Quantum services
for this work; the views expressed are those of the author, and do not
reflect the official policy or position of IBM or the IBM Quantum team.
Claude Opus 5.5 (Anthropic) assisted with numerical code,
figures and text, and Gemini (Google) with critical reading; the author
directed the work, verified the results, and is solely responsible for
the content.
\end{acknowledgments}

\bibliography{FractalPeriodLocking_v2}

\clearpage
\pagestyle{empty}
\AddToShipoutPictureFG{\ifnum\value{page}>7 \AtPageLowerLeft{\includegraphics[page=\the\numexpr\value{page}-7\relax]{FractalPeriodLocking_v2_SM.pdf}}\fi}
\makeatletter
\@tempcnta=0
\loop\ifnum\@tempcnta<10
\advance\@tempcnta by 1
\thispagestyle{empty}\null\clearpage
\repeat
\makeatother
\end{document}